\documentclass[a4paper]{PoS}

\DeclareMathAlphabet{\mathcal}{OMS}{cmsy}{m}{n}
\newcommand{\adeg}{\ensuremath{^\circ}}

\newcommand{\etal}{{et al.}}

\title{Results from monitoring TeV blazars with HAWC}

\ShortTitle{Monitoring TeV blazars with HAWC}

\author{\speaker{Robert J. Lauer} $^{a}$ and Patrick W. Younk $^{b}$ for the HAWC Collaboration $^{c}$\\
        \llap{$^a$}
        University of New Mexcio, Department of Physics and Astronomy, Albuquerque, NM, USA\\
        \llap{$^b$}
        Physics Division, Los Alamos National Laboratory, Los Alamos, NM, USA\\
        \llap{$^c$}
        For a complete author list, see \href{http://www.hawc-observatory.org/collaboration/icrc2015.php}{www.hawc-observatory.org/collaboration/icrc2015.php}.\\
        E-mail: \email{rlauer@phys.unm.edu}
        }
\abstract{The recently completed High Altitude Water Cherenkov (HAWC) gamma-ray observatory has been taking data with a partial array for more than one year and is now operating with >95\% duty cycle in its full configuration. With an instantaneous field of view of 2 sr, two-thirds of the sky is surveyed every day at gamma-ray energies between approximately 100 GeV and 100 TeV. Any source location in the field of view can be monitored each day, with an exposure of up to $\sim6$ hours.
These unprecedented observational capabilities allow us to continuously scan the highly variable extra-galactic gamma-ray sky. 
By monitoring the flaring behavior of Active Galactic Nuclei we aim to significantly increase the observational data base for characterizing particle acceleration mechanisms in these sources and for studying cosmological properties like the extra-galactic background light. 
In this work we present first studies of data taken between June 2013 and July 2014 with a partial array configuration. Flux light curves, binned in week-long intervals, for the TeV-emitting blazars Markarian 421 and 501 are discussed with respect to indications of flaring states and we highlight coincident multi-wavelength observations. Results for both sources show indications of gamma-ray flare observations and demonstrate that a water Cherenkov detector can monitor TeV-scale variability of extra-galactic sources on weekly time scales. The analysis methods presented here can provide daily flux measurements with a minimum time interval of one transit and will be applied to new data from the completed HAWC array for monitoring of blazars and other transients.}

\FullConference{The 34th International Cosmic Ray Conference,\\
		30 July- 6 August, 2015\\
		The Hague, The Netherlands}

\begin{document}

\section{Introduction}

Active Galactic Nuclei (AGN) are the most populated source class of very high energy (VHE; $>$ 100 GeV) gamma-ray emission.~\footnote{See list at http://tevcat.uchicago.edu .} Those AGN with jets oriented close to the line of sight are called blazars and among them are some of the brightest VHE sources. VHE fluxes from blazars are often highly variable, and periods of increased activity, called flares from here on, have been observed with fluxes increasing by an order of magnitude~\cite{bib:punch} and with variability on time scales from months down to minutes~\cite{bib:aharonian2007,bib:albert2007}. 
It is poorly understood where along the jet these gamma rays are emitted and if the underlying processes that accelerate particles to TeV or higher energies are dominated by populations of leptonic or hadronic particles. Only a more detailed coverage of flares over a wide range of wavelengths can constrain such models as in~\cite{bib:boettcher}.  
Most of the observations at VHE energies have been performed with Imaging Air Cherenkov Telescopes (IACTs) that can only operate during clear, dark nights and normally monitor only one source in the field of view at any time. 
Their capacity for long term studies of the frequency of blazar flaring states is therefore limited. 
Furthermore, extended monitoring that is not biased by following-up on flaring alerts at other wavelengths can help to study the possibility of flares limited to TeV energies, so-called orphan flares, which might help to identify hadronic acceleration processes~\cite{bib:kraw}. 
Studies of the flux variability in AGN and their underlying acceleration mechanisms will greatly profit from unbiased monitoring observations with high duty cycles that are made possible with the HAWC Observatory. In addition, bright flares offer a unique opportunity to constrain or measure the extra-galactic background light~\cite{bib:stecker} and inter-galactic magnetic fields~\cite{bib:neronov}.

\section{Analysis method}

\subsection{The HAWC Observatory}
\label{section-det}

The \textbf{H}igh \textbf{A}ltitude \textbf{W}ater \textbf{C}herenkov (HAWC) Observatory is located at an altitude of 4,100~m above sea level on the slope of the Sierra Negra volcano ($19.0\adeg$N $97.3\adeg$W) in the state of Puebla, Mexico. The detector is optimized for measuring extensive air showers induced by gamma rays with energies between approximately 100~GeV and 100~TeV. Completed in March 2015, HAWC is an array of 300 Water Cherenkov Detectors (WCDs).
Each WCD is composed of a large steel tank holding $\sim200,000$ liters of purified water in a light-proof bladder. Four photo-multiplier tubes (PMTs), located at the bottom, can detect the Cherenkov light from the particles in an air shower passing through the array.
By analyzing the recorded light intensity distribution as a proxy for the particle density and fitting the plane of the shower front from the photon arrival times it is possible to reconstruct the size and incident direction of an air shower event.
While most of the air showers recorded are produced by hadronic primaries, this background can be significantly reduced by identifying large charge deposits outside the core regions indicative of muons, which are not expected in gamma-ray showers. More details about the reconstruction, data taking and hadron rejection are discussed elsewhere in these proceedings~\cite{bib:icrcPretz,bib:icrcSmith}.
\\
The modular design of HAWC made it possible to start continuous scientific operation before the completion of the whole detector. 
The results presented here are from data taken between June 13, 2013, when HAWC was operating with 307 active PMTs in 90 WCDs, and July 9, 2014, by which time the array had grown to include 495 active PMTs in 135 WCDs.
During the whole period the detector was operating with a duty cycle $\sim90$~\% with down time mostly due to infrequent maintenance activities that required a complete shutdown.

\subsection{Data reduction}

After calibration and reconstruction, the air shower events are sorted into 10 analysis bin, primarily defined by the fraction of PMTs that measured a signal in the event. The lower threshold of the first analysis bin is 4.5\% of all active channels, thus requiring $\sim20$ PMTs with photon hits in the same event. 
As a measurement of shower size, the number of PMTs with signals serves both as a rough proxy of the energy of the primary particle and as an event quality parameter, since larger events with more signals will on average have a more precise angular reconstruction. For each analysis bin, additional cuts on several event parameters reduce the number of hadronic background events, as mentioned in section~\ref{section-det}.
\\
Due to the increasing absorption of secondary particles in the atmosphere, the sensitivity to gamma rays is a function of zenith angle. For the time-dependent analysis in this contribution we use the duration of one source transit over HAWC as the smallest time interval under consideration, thus simplifying this dependence to be a function of the source declination. A HAWC analysis targeted for blazars flares with durations of less than one transit is described elsewhere in these proceedings~\cite{bib:icrcTom}. Since the gamma-ray efficiency of HAWC for zenith angles larger than $45^{\circ}$ is negligible, a source transit can be considered to last at most $\sim6$~hours.
In order to store the data in time intervals that cover only one transit, the reconstructed events are sorted into sidereal days, starting at midnight local sidereal time at the HAWC longitude of $97.3\adeg$W. This choice leads to transits being split in two for sources with right ascension (RA) $<3$~h or $>21$~h, which are analyzed with a separate binning of the data starting at noon sidereal time. 
\\
For each sidereal day and each of the ten analysis bins, a sky map of event counts is produced by populating pixels with an average spacing of $\sim0.1^{\circ}$. In general, these maps are still dominated by hadronic background events, and direct integration~\cite{bib:atkins}, averaging counts over two hours around any location, yields an estimate of this background. 
The estimated background counts in each pixel are stored in a second map with the same grid structure, so that a difference of the two maps provides event excess counts.

\subsection{Likelihood analysis}

To analyze the combination of the 10 analysis bins for a particular region of the sky we employ a likelihood ratio test, a standard technique in particle physics~\cite{bib:PDG}.
The software implementation for performing this test on a given set of HAWC data maps interfaces directly with the general HAWC software framework and is called \textbf{Li}kelihood \textbf{F}itting \textbf{F}ramework (LiFF), described in detail elsewhere in these proceedings~\cite{bib:liff}.
In this contribution, the only signal model considered represents a gama-ray point source with the differential flux energy spectrum described by a power law with index $s$ and an optional exponential cut-off $c$:
\begin{equation}
\label{eq:spectrum}
 \frac{dN_{\mbox{\tiny ph}}}{dE} = F_0 \cdot \left(\frac{E}{\mbox{TeV}}\right)^{-s} \cdot \exp{\left(-\frac{E}{c}\right)}\quad.
\end{equation}
In LiFF, this input flux is convoluted with a detector response function that includes the point spread function and efficiency of triggers and cuts, depending on primary energy and incident angle. For one source transit over HAWC, the signal hypothesis contributions as a function of zenith angle are summed and yield the expected number of events $S_{b,p}$ per analysis bin $b$ and pixel $p$ in the map
for one sidereal day.
In cases were the coverage of a source transits is interrupted, for example due to a detector shutdown, the lost signal fraction compared to a full transit is calculated by excluding the gap period from the integration over zenith angle and this fraction is then used to reduce the expected event count accordingly.
The likelihood of a hypothesis defined by $\{F_0,s,c\}$ given an observation $\vec{N}$ of numbers of events in all bins and pixels is expressed via the likelihood function  
\begin{equation}
\label{eq:llh}
 \mathcal{L}_S \left( \{F_0,s,c\}|\vec{N}\right) = \prod_b{\prod_p{ P(N_{b,p},\lambda_{b,p})}} \quad ,
\end{equation}
where $P(N_{b,p},\lambda_{b,p})$ is the Poisson distribution for a mean expectation $\lambda_{b,p} = S_{b,p}+B_{b,p}$, i.e. the sum of the expected signal and the number of background events estimated from data for analysis bin $b$ and pixel $p$.
In the likelihood ratio test, the result of eq.~(\ref{eq:llh}) is divided by the likelihood value $\mathcal{L}_B$ for a background-only assumption ($S_{b,p} = 0$). 
In the software implementation, this ratio is expressed as a difference $\Delta\ln{\mathcal{L}}$ of the logarithms of the two likelihood values and the test statistic is defined as $TS = 2 \Delta\ln{\mathcal{L}}$ .                                                                                                                                                                                                                                                                                                                                                                                                                                                                                                                                                                                                                                                                                                                                                                      
$TS$ is then numerically maximized by iteratively changing the input parameters, yielding those values that have the highest likelihood of describing the observed data for the point source model assumption.
\\
In the analysis presented here, the spectral index $s$ and cut-off value $c$ in eq.~(\ref{eq:spectrum}) were kept constant for each target source and only the normalization $F_0$ was left free in the likelihood fit. 
This is reasonable since even the one-year time-integrated data has a limited sensitivity to spectral parameters due to small statistics. A first look at AGN spectra with HAWC is discussed in~\cite{bib:icrcSara}.

\subsection{Monitoring strategy}
\label{monitoring}
Given the limited sensitivity of the partial HAWC detector, the results of a likelihood fit to the data from one transit exhibits large uncertainties in this early analysis. The Crab Nebula, the brightest steady source seen by HAWC, yields an average significance of less than 1~$\sigma$ per transit. 
For this reason, the monitoring presented in this contribution is performed for time intervals of 7~transits, yielding a total of $n_t = 56$ light curve bins. 
It is therefore insensitive to flux variability on shorter time scales, but provides more stable result for a first analysis of the blazar monitoring potential of HAWC. 
A light curve for the Crab Nebula, derived with the methods described here, is shown elsewhere in these proceedings~\cite{bib:icrccrab} and is consistent with a constant flux over time.
\\
The integrated life time of the data in the period June 13, 2013, to July 9, 2014, is equivalent to approximately 350 transits, varying slightly ($\sim5$ days) for different source right ascensions due to small variations in the exposure. Among the locations that surpass a pre-trial significance in the time-integrated analysis, only two are identified as known extra-galactic sources, the two VHE-emitting blazars Markarian 421 and Markarian 501. 
The monitoring results for these two sources are discussed in the following section.
\\
The Crab Units on the right axes in the light curve figures in section~\ref{results} are defined by dividing the measured flux values by the average flux of the Crab nebula as observed in HAWC, derived by assuming a simple power law spectrum with $s=2.63$~\cite{bib:hesscrab} and the flux normalization result for the data for the whole period. In general, we estimate the systematic uncertainty of the HAWC fluxes to be $\sim40\%$ which is expected to be reduced with further progress on the data analysis~\cite{bib:icrccrab}.

\section{Results}
\label{results}

\subsection{Markarian 421}

\begin{figure}[t]
\includegraphics[width=.99\textwidth]{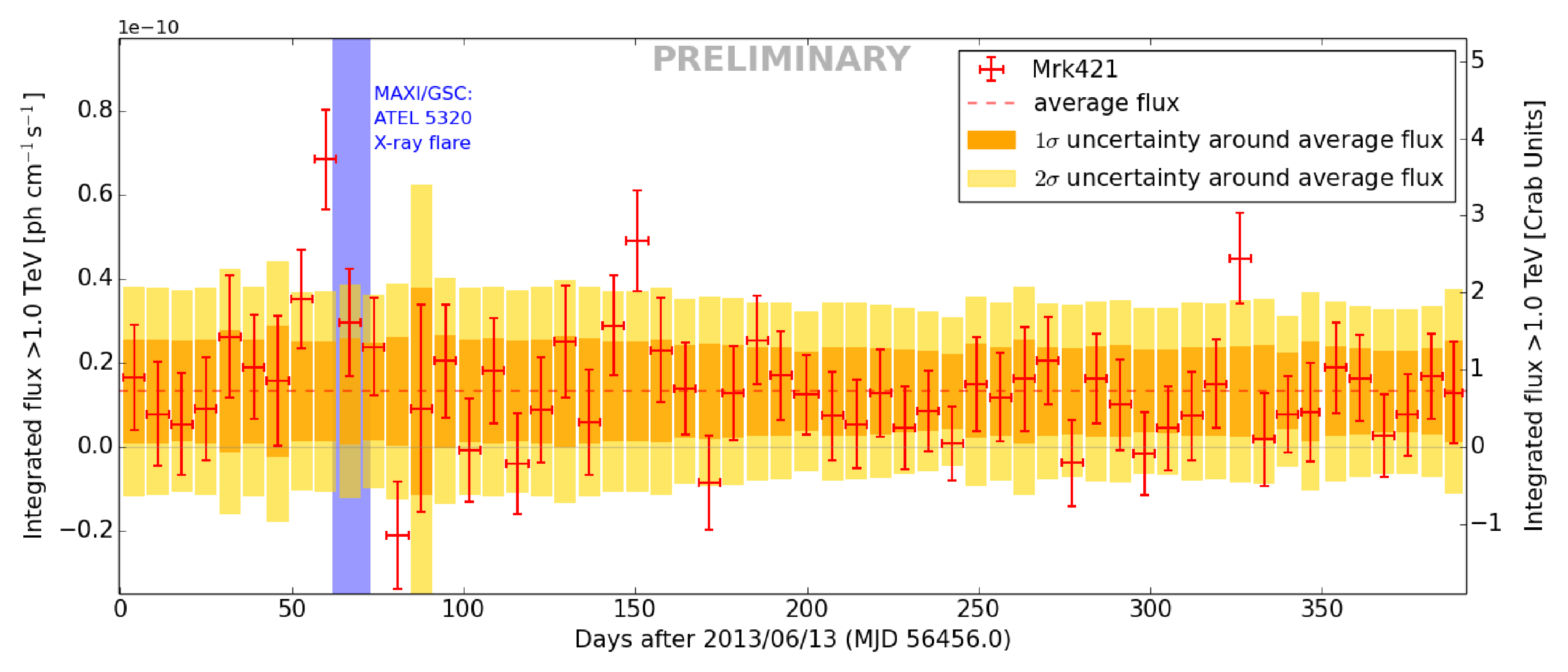}
\caption{Flux light curve of Markarian 421 for the time between June 13, 2013, and July 9, 2014, in intervals of 7 days, with horizontal error bars indicating the time span between the start of the first and the end of the last transit over HAWC. 
Integrated fluxes above 1 TeV for a differential spectrum with power law index 2.2 and exponential cut-off at 5~TeV are shown in photons per cm$^2$ s$^2$ on the left axis and divided by the average Crab flux observed by HAWC on the right axis. The statistical 1 and 2~$\sigma$ flux uncertainties of the individual measurements are displayed as orange/yellow bars around the weighted average flux (dashed line).
The large uncertainties in the flux around day $\sim90$ are due to a period of construction and maintenance in September 2013 during which HAWC was shut down during day time for several days in a row, creating large gaps in the transit coverage.
The blue band indicates the period over which a strong increase in X-ray flux was observed by MAXI~\cite{bib:atel5320}.}
\label{fig:mrk421}
\end{figure}

Markarian 421 (Mrk 421) is a BL Lacertae type blazar with a redshift of $z\approx 0.03$ that has been well studied by many VHE gamma-ray observatories and is known to exhibit a high degree of variability~\cite{bib:tluczykont}. For this analysis, we use a simple power law signal hypothesis with index $s=2.2$ and exponential cut-off at $c=5$~TeV, based on typical established values, for example~\cite{bib:krennrich,bib:veritas421}.
The results from the likelihood fit of flux normalizations in each time interval are displayed as integrated fluxes in the light curve shown in Fig.~\ref{fig:mrk421}.
\\
For the period considered here, the data base of The Astronomer's Telegram~\footnote{http://www.astronomerstelegram.org} lists only one report of increased activity in Mrk 421, an X-ray observation at 2-4~keV by the MAXI/GSC instrument during August 2013~\cite{bib:atel5320}. 
The corresponding blue band in Fig.~\ref{fig:mrk421} roughly visualizes the length of that period of X-ray brightening.
The highest flux value obtained in the HAWC analysis coincides with the onset of the MAXI observation, several days before their reported peak X-ray flux at Modified Julian Date (MJD) 56524.
The chance probability of the HAWC measurment alone to occur as a fluctuation, based on the $\chi^2$ value of the excess compared to a constant average flux and a trial factor for 56 time bins, is $1.4\cdot10^{-5}$ ($4.3\sigma$).
This is a strong indication that both HAWC and MAXI data show the same flaring event, though we have not yet developed a statistical method to determine the significance of this multi-wavelength correlation.

\subsection{Markarian 501}

\begin{figure}[t]
\includegraphics[width=.99\textwidth]{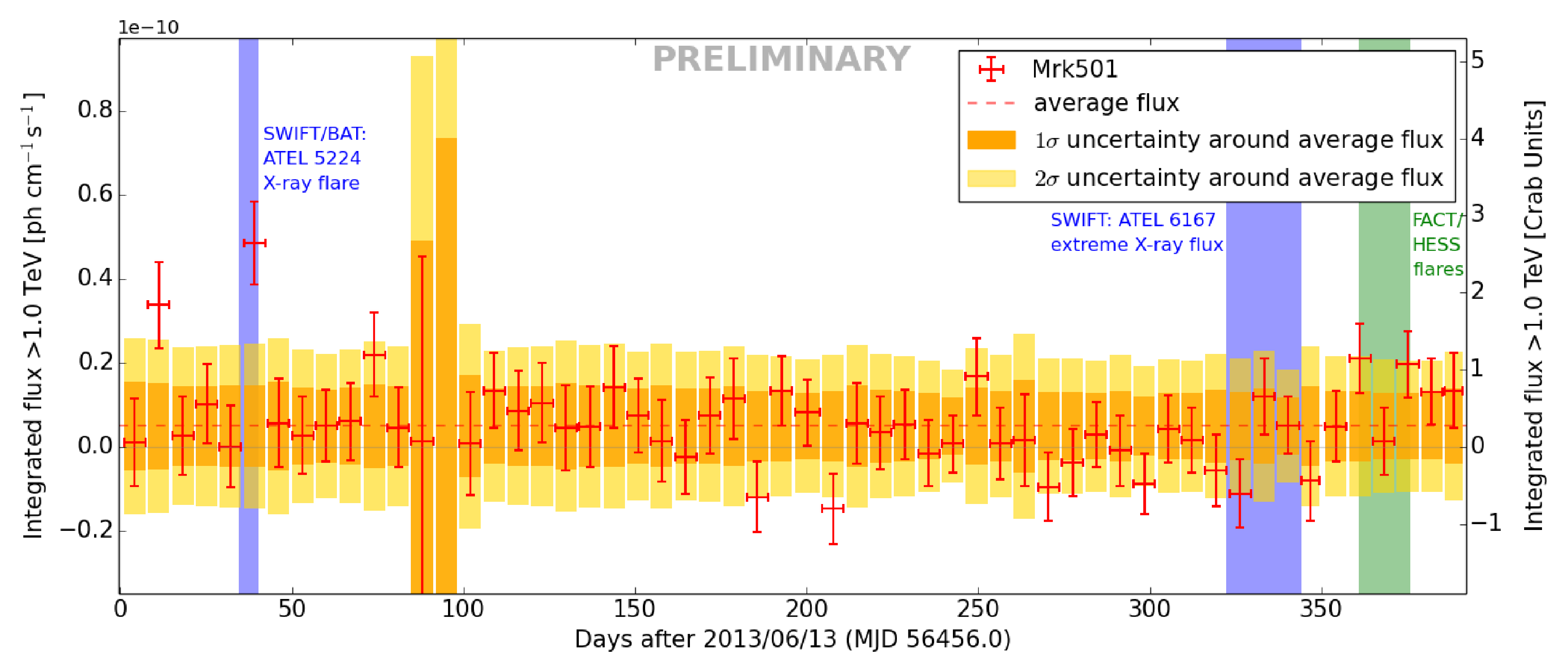}
\caption{Flux light curve of Markarian 501 for the time between June 13, 2013, and July 9, 2014, in intervals of 7 days, with horizontal error bars indicating the time span between the start of the first and the end of the last transit over HAWC. Integrated fluxes above 1 TeV for a differential spectrum with power law index 2.8 are shown in photons per cm$^2$ s$^2$ on the left axis and divided by the average Crab flux observed by HAWC on the right axis. 
The statistical 1 and 2~$\sigma$ flux uncertainties of the individual measurements are displayed as orange/yellow bars around the weighted average flux (dashed line). 
The large uncertainties in the flux around day $\sim90$ are due to a period of construction and maintenance in September 2013 during which HAWC was shut down during day time for several days in a row, creating large gaps in the transit coverage.
The blue/green bands indicating increased X-ray/gamma-ray fluxes observed by other instruments are based on publicly available data and discussed and referenced in the text.}
\label{fig:mrk501}
\end{figure}

Markarian 501 (Mrk 501) is another BL Lacertae type blazar with a redshift of $z\approx 0.03$ and with a history of known VHE gamma-ray flares, see for example~\cite{bib:tluczykont}. For the HAWC analysis, we used a fixed spectral shape with index $s=2.8$ and no cut-off ($c=\infty$) in eq.~\ref{eq:spectrum}, corresponding to typical values~\cite{bib:hegra501}. The HAWC flux light curve results are shown in Fig.~\ref{fig:mrk501}. 
The highest flux value in the HAWC analysis is observed during a week in July 2013 that largely overlaps with a period of increased flux observed by the Swift/BAT hard X-ray transient monitor in the 15-50~keV band, reported in~\cite{bib:atel5224}. The left vertical blue band visualizes the approximate extent of the X-ray observation that revealed a peak flux at MJD 56495, in the right half of the band. Our a-posteriori analysis cannot quantify the probability for this coincidence but seems to indicate that the HAWC observation shows the TeV counterpart of the X-ray flare. 
The chance probability of the largest HAWC flux measurment to occur as a fluctuation above the constant average flux, considering a trial factor for 56 time bins, is $1.9\cdot10^{-4}$ ($3.7\sigma$).
A second X-ray flaring state was observed in May 2014 in the lower 0.3-10 keV energy band by the X-ray Telescope (XRT) onboard the Swift satellite~\cite{bib:atel6167} and is included in Fig.~\ref{fig:mrk501} as the vertical blue band on the right side. No clear flux feature is observed in the HAWC data around that time.
\\
The green vertical band indicates a period where the VHE gamma-ray telescope FACT observed several flares from Mrk 501 with flux variations changing on the time scale of hours~\cite{bib:fact}. 
This triggered a follow-up observation by the VHE gamma-ray telescope H.E.S.S.~\cite{bib:atel6268}, with a peak flux at MJD 56832. None of the weekly HAWC flux measurments during that period are larger than 2$\sigma$ above the flux average. 
It is important to note the transits time windows of the HAWC observations never overlap with the night time observations of either FACT or H.E.S.S. and that the averaging over 7~days in the HAWC analysis reduces the sensitivity to flux variability over less than one hour reported for this flare. 
The HAWC results presented here reveal that the week-averaged fluxes during these three weeks were not significantly higher than the average value over the whole year.

\section{Conclusion}

Data taken between June 2013 and July 2014 with an early configuration of the HAWC Observatory, operating with approximately one third of the completed array, have been analyzed with a likelihood technique to produce flux measurments for 56 week-long time intervals. These light curves for the two clearly detected extra-galactic sources, BL Lacertae objects Markarian 421 and Markarian 501, both show indications of variability and both have the highest VHE flux values coinciding with X-ray flare observations. Detailed joint analyses of the multi-wavelength data and a finer time binning for selected periods of interest could reveal more features of the VHE gamma-ray flux during known flares of these sources.
\\
New HAWC data with improved sensitivity from several months of operating a grown configuration and, after March 2015, the completed HAWC Observatory are now available for analysis. With the increased sensitivity of the larger array, a light curve analysis with reduced flux uncertainties and shorter interval lengths of 1 transit will be possible and can provide regular monitoring for any blazar, or in fact any sky location, in the field of view of HAWC.

\section*{Acknowledgments}
\footnotesize{
We acknowledge the support from: the US National Science Foundation (NSF);
the US Department of Energy Office of High-Energy Physics;
the Laboratory Directed Research and Development (LDRD) program of
Los Alamos National Laboratory; Consejo Nacional de Ciencia y Tecnolog\'{\i}a (CONACyT),
Mexico (grants 260378, 55155, 105666, 122331, 132197, 167281);
Red de F\'{\i}sica de Altas Energ\'{\i}as, Mexico;
DGAPA-UNAM (grants IG100414-3, IN108713,  IN121309, IN115409, IN113612);
VIEP-BUAP (grant 161-EXC-2011);
the University of Wisconsin Alumni Research Foundation;
the Institute of Geophysics, Planetary Physics, and Signatures at Los Alamos National Laboratory;
the Luc Binette Foundation UNAM Postdoctoral Fellowship program.
}


\begin{thebibliography}{99}
\bibitem{bib:punch} Punch, M. \etal, \textit{Detection of TeV photons from the active galaxy Markarian 421.} Nature,  \textbf{358} (1992) 477.
\bibitem{bib:aharonian2007} Aharonian, F. \etal, \textit{An Exceptional Very High Energy Gamma-Ray Flare of PKS 2155-304.} Astrophys. J. , \textbf{664} (2007) L71.
\bibitem{bib:albert2007} Albert, J. \etal, \textit{Variable Very High Energy $\gamma$-Ray Emission from Markarian 501.} Astrophys. J., \textbf{669} (2007) 862.
\bibitem{bib:boettcher} B\"ottcher, M. \etal, \textit{Leptonic and Hadronic Modeling of 111 Fermi-detected Blazars.} Astrophys. J. \textbf{768} (2013) 54.
\bibitem{bib:kraw} Krawczynski, H. \etal, \textit{Multiwavelength Observations of Strong Flares From the TeV-Blazar 1ES 1959+650.} Astrophys. J., \textbf{601} (2004) 151.
\bibitem{bib:stecker} Stecker, F. \etal, \textit{TeV gamma rays from 3C 279: A possible probe of origin and intergalactic infrared radiation fields.} Astrophys. J. \textbf{390} (1992) L49.
\bibitem{bib:neronov} Neronov, A. \etal, \textit{A method of measurement of extragalactic magnetic fields by TeV gamma ray telescopes.} JETP Lett. \textbf{85} (2007) 473.
\bibitem{bib:icrcPretz} HAWC Collaboration, Pretz, J., \textit{Highlights from the High Altitude Water Cherenkov Observatory.} Proc. 34th ICRC (2015), The Hague, The Netherlands
\bibitem{bib:icrcSmith} HAWC Collaboration, Smith, A., \textit{HAWC: Design, Operation, Reconstruction and Analysis.} Proc. 34th ICRC (2015), The Hague, The Netherlands
\bibitem{bib:icrcTom} HAWC Collaboration, Weisgarber, T. \etal, \textit{Blazar Alerts with the HAWC Online Flare Monitor.} Proc. 34th ICRC (2015), The Hague, The Netherlands
\bibitem{bib:atkins} Atkins, R. \etal, \textit{Observation of TeV Gamma Rays from the Crab Nebula with Milagro Using a New Background Rejection Technique.} Astrophys. J., \textbf{595} (2003) 803
\bibitem{bib:PDG} Beringer, J. \etal, \textit{Review of Particle Physics.} Phys. Rev. D \textbf{86} (2012) 010001.
\bibitem{bib:liff}  HAWC Collaboration, Younk, P. \etal, \textit{A high-level analysis framework for HAWC.} Proc. 34th ICRC (2015), The Hague, The Netherlands. 
\bibitem{bib:icrcSara}  HAWC Collaboration, Couti\~no, S. \etal, \textit{Selection of AGN to study the extragalactic background light with HAWC.} Proc. 34th ICRC (2015), The Hague, The Netherlands.
\bibitem{bib:icrccrab} HAWC Collaboration, Salesa Greus, F. \textit{Observations of the Crab Nebula with Early HAWC Data.} Proc. 34th ICRC (2015), The Hague, The Netherlands.
\bibitem{bib:hesscrab} Aharonian, F. \etal, \textit{Observations of the Crab Nebula with H.E.S.S.} Astron. Astrophys. \textbf{457} (2006) 899.
\bibitem{bib:tluczykont} Tluczykont, M. \etal, \textit{Long-term lightcurves from combined unified very high energy $\gamma$-ray data.} Astron. Astrophys. \textbf{524} (2010) A48.
\bibitem{bib:krennrich} Krennrich, F. \etal, \textit{Discovery of spectral variability of Markarian 421 at TeV energies.} Astrophys. J. \textbf{575} (2002) L9.
\bibitem{bib:veritas421} Acciaria, V. \etal, \textit{Observation of Markarian 421 in TeV gamma rays over a 14-year time span.} Astrop. Phys. \textbf{54} (2014) 1.
\bibitem{bib:hegra501} Aharonian, F. \etal, \textit{The TeV Energy Spectrum of Markarian 501 Measured with the Stereoscopic Telescope System of HEGRA during 1998 and 1999.} Astrophys. J. \textbf{546} (2001) 898.
\bibitem{bib:fact} Anderhub, H. \etal, \textit{Design and operation of FACT -- the first G-APD Cherenkov telescop.} JINST \textbf{8} (2013) P06008. \texttt{http://fact-project.org/monitoring.}
\bibitem{bib:atel5320} Ueno, S. \etal, \textit{MAXI/GSC detection of a bright X-ray flare from Mrk 421.} ATel \textbf{5320} (2014).
\bibitem{bib:atel5224} Krimm, H. \etal \textit{Swift/BAT detects a new outburst of the blazar Mrk 501.} ATel \textbf{5224} (2014).
\bibitem{bib:atel6167} Kapanadze, B., \textit{The Highest Historical 0.3-10 keV Flux in HBL Source Markarian 501.} ATel \textbf{6167} (2014).
\bibitem{bib:atel6268} Stegmann, C., \textit{Increased VHE activity from Mrk 501 detected with H.E.S.S.} ATel \textbf{6268} (2014).
\end{thebibliography}
\end{document}